\documentclass[reprint,superscriptaddress,nofootinbib,amsmath,amssymb,aps,prl,twocolumn]{revtex4-2}
\usepackage{float}
\usepackage{graphicx}
\usepackage{dcolumn}
\usepackage{bm}
\usepackage[colorlinks]{hyperref}
\usepackage{csquotes}
\usepackage[T1]{fontenc}
\usepackage{lmodern} 
\usepackage{lipsum}
\usepackage{siunitx}

\newcommand{\bra}[1]{\left< #1 \right\vert}
\newcommand{\ket}[1]{\left\vert #1 \right>}

\newcommand{\abs}[1]{\left\vert #1 \right\vert}

\newcommand*\chem[1]{\ensuremath{\mathrm{#1}}}

\begin{document}

\preprint{APS/123-QED}

\title{Conditional Quantum Plasmonic Sensing}

\author{Fatemeh Mostafavi}
\email{fmosta1@lsu.edu}
\affiliation{Quantum Photonics Laboratory, Department of Physics \& Astronomy, Louisiana State University, Baton Rouge, LA 70803, USA}

\author{Zeinab Jafari}
\affiliation{School of Engineering and Sciences, Tecnol\'ogico de Monterrey, Monterrey, Nuevo Leon 64849, Mexico}

\author{Michelle L. J. Lollie}
\affiliation{Quantum Photonics Laboratory, Department of Physics \& Astronomy, Louisiana State University, Baton Rouge, LA 70803, USA}

\author{Chenglong You}
\affiliation{Quantum Photonics Laboratory, Department of Physics \& Astronomy, Louisiana State University, Baton Rouge, LA 70803, USA}

\author{Israel De Leon}
\affiliation{School of Engineering and Sciences, Tecnol\'ogico de Monterrey, Monterrey, Nuevo Leon 64849, Mexico}

\author{Omar S. Maga\~na-Loaiza}
\affiliation{Quantum Photonics Laboratory, Department of Physics \& Astronomy, Louisiana State University, Baton Rouge, LA 70803, USA}

\date{\today}

\begin{abstract}

The possibility of using weak optical signals to perform sensing of delicate samples constitutes one of the main goals of quantum photonic sensing. Furthermore, the nanoscale confinement of electromagnetic near fields in photonic platforms through surface plasmon polaritons has motivated the development of highly sensitive quantum plasmonic sensors. Despite the enormous potential of plasmonic platforms for sensing, this class of sensors is ultimately limited by the quantum statistical fluctuations of surface plasmons. Indeed, the fluctuations of the electromagnetic field severely limit the performance of quantum plasmonic sensing platforms in which delicate samples are characterized using weak near-field signals. Furthermore, the inherent losses associated with plasmonic fields levy additional constraints that challenge the realization of sensitivities beyond the shot-noise limit. Here, we introduce a protocol for quantum plasmonic sensing based on the conditional detection of plasmons. We demonstrate that the conditional detection of plasmonic fields, via plasmon subtraction, provides a new degree of freedom to control quantum fluctuations of plasmonic fields. This mechanism enables improvement of the signal-to-noise ratio of photonic sensors relying on plasmonic signals that are comparable to their associated field fluctuations. Consequently, the possibility of using weak plasmonic signals to sense delicate samples, while preserving the sample properties, has important implications for molecule sensing, and chemical detection
.
\end{abstract}

\maketitle
The possibility of controlling the confinement of plasmonic near-fields at the subwavelength scale has motivated the development of a variety of extremely sensitive nanosensors \cite{ NanoToday, Hwang, Maier, Lee1}. Remarkably, this class of sensors offers unique resolution and sensitivity properties that cannot be achieved through conventional photonic platforms in free space \cite{ChenglongAIP,Slussarenko2017,Lee1, Polino}. In recent decades, the fabrication of metallic nanostructures has enabled the engineering of surface plasmon resonances to implement ultrasensitive optical transducers for detection of various substances ranging from gases to biochemical species \cite{NanoToday,Hwang,Lee1}. Additionally, the identification of the quantum mechanical properties of plasmonic near-fields has prompted research devoted to exploring mechanisms that boost the sensitivity of plasmonic sensors \cite {Altewischer, Akimov,  Tame2013, chenglongreview, chenglongnature}. 

The scattering paths provided by plasmonic near-fields have enabled robust control of quantum dynamics \cite{BenjaminVest,Schouten,maganaloaizaexotic2016, chenglongnature}. Indeed, the additional degree of freedom provided by plasmonic fields has been used to harness the quantum correlations and quantum coherence of photonic systems \cite{Alexander,chenglongnature,Dongfang,Schouten}. Similarly, this exquisite degree of control made possible the preparation of plasmonic systems in entangled and squeezed states \cite{Vest,Huck,Fasel,Martino}. Among the large variety of quantum states that can be engineered in plasmonic platforms \cite{Tame2013,chenglongreview}, entangled systems in the form of N00N states or in diverse forms of squeezed states have been  used to develop quantum sensors \cite{Heeres2013,Chen:18,Lee1,Dowran2018,Lee2016,Dowran2018}. In principle, the sensitivity of these sensors is not constrained by the quantum fluctuations of the electromagnetic field that establish the shot-noise limit \cite{lee,Polino}. However, due to inherent losses of plasmonic platforms, it is challenging to achieve sensitivities beyond the shot-noise limit under realistic conditions \cite{ChenglongAIP}. Despite existing obstacles, recent work demonstrates the potential of exploiting nonclassical properties of plasmons to develop quantum plasmonic sensors for detection of antibody complexes, single molecules, and to perform spectroscopy of biochemical substances \cite{Salazar,Kongsuwan,alashnikov,Mauranyapin2017}. 

\begin{figure*}[!ht]
\centering
\includegraphics[width=0.95\textwidth]{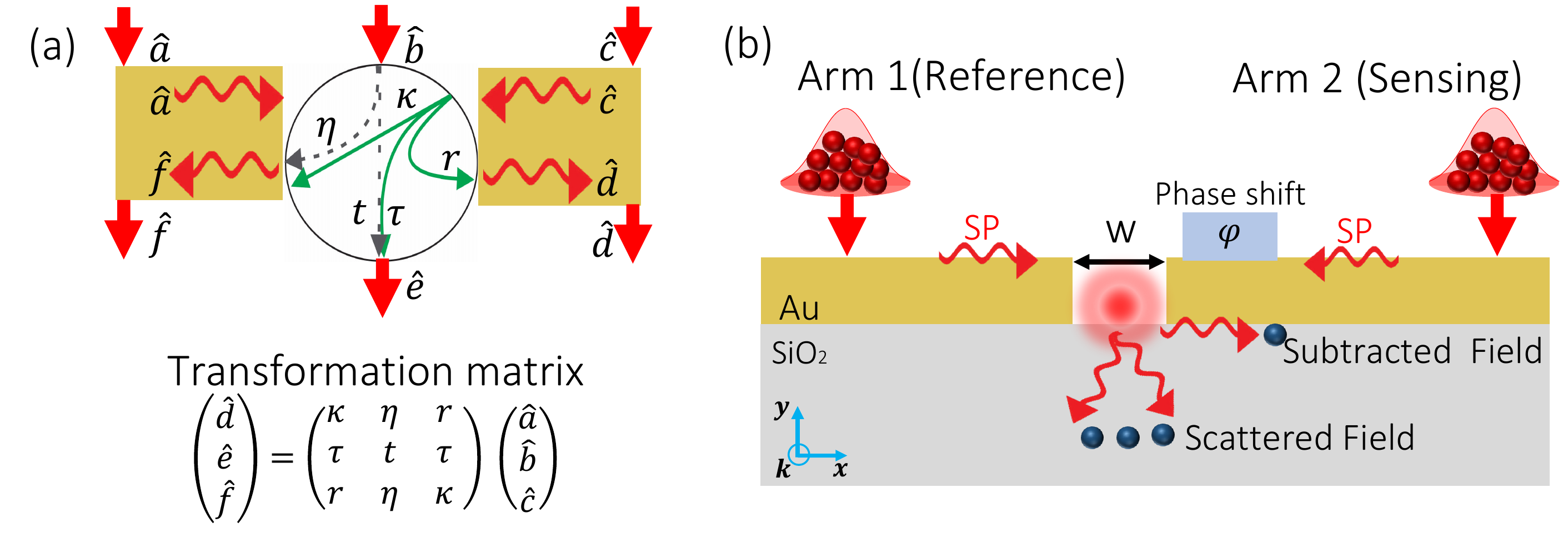}
  \caption{(a) Schematic diagram of the interactions in a plasmonic nanoslit. The plasmonic nanostructure has three input and three output ports. The photonic mode at the input is described by the operator $\hat{b}$, whereas the two plasmonic modes are represented by $\hat{a}$ and $\hat{c}$. These modes are coupled to the plasmonic modes $\hat{d}$ and $\hat{f}$, and to the photonic mode $\hat{e}$ at the output of the nanostructure. As described by the transformation matrix, the parameters $\kappa$, $\tau$, $r$, $\eta$, and $t$ represent the coupling coefficients among the ports of the nanostructure. For sake of clarity, the diagram only illustrates the coupling paths for the input modes $\hat{b}$ and $\hat{c}$. The diagram in (b) shows the design of our simulated plasmonic sensor, comprising a slit of width w in a $200$ nm gold thin film. Here, the plasmonic structure is illuminated by two thermal multiphoton sources that excite two plasmonic fields with super-Poissonian statistics (the input grating couplers are not shown in the figure). The two counter-propagating surface plasmon (SP) modes  interfere at the interface between the gold layer and the \chem{SiO_2} substrate. The interference conditions are defined by the phase shift $\varphi$ induced in one of the plasmonic modes by the substance that we aim to sense. 
  }
   \label{fig:figure1}
\end{figure*}  

Here, we explore a new scheme for quantum sensing based on plasmon-subtracted thermal states \cite{Dakna, Loaiza2019, HashemiRafsanjani:17}. Our work offers an alternative to quantum sensing protocols relying on entangled or squeezed plasmonic systems \cite{Vest,Huck,Fasel,Martino,Heeres2013,Chen:18,Lee1,Dowran2018,Lee2016,Dowran2018}.  We use a sensing architecture based on a nanoslit plasmonic interferometer \cite{biosensordeleon}. It provides a direct relationship between the light exiting the interferometer and the phase shift induced in one of its arms by the substance to be sensed (analyte). We introduce a conditional quantum measurement on the interfering plasmonic fields via the subtraction of plasmons. We show that this process enables the reduction of quantum fluctuations of the sensing field and increases the mean occupation number of the plasmonic sensing platform\cite{Loaiza2019, HashemiRafsanjani:17}. Furthermore, plasmon subtraction provides a method for manipulating the signal- to-noise ratio (SNR) associated with the measurement of phase shifts. 
We demonstrate that the reduced fluctuations of plasmonic fields leads to an enhancement in the estimation of a phase shift. The performance of our protocol is quantified through the uncertainty associated to phase measurements. We point out that the reduced uncertainties in the measurement of phases leads to better sensitivities of our sensing protocol. This study is conducted through a quantum mechanical model that considers the realistic losses that characterize a plasmonic nanoslit sensor. We report the probabilities of successfully implementing our protocol given the occupation number of the plasmonic sensing fields and the losses of the nanostructure. Our analysis suggests that our protocol offers practical benefits for lossy plasmonic sensors relying on  weak near-field signals \cite{Maga_a_Loaiza_2016}.  Consequently, our platform can have important implications for plasmonic sensing of delicate samples such as molecules, chemical substances or, in general, photosensitve materials \cite{Salazar,Kongsuwan,alashnikov,Mauranyapin2017}. 
  
 We first discuss the theoretical model that we use to describe conditional quantum measurements applied to a thermal plasmonic system. Fig. \ref{fig:figure1}a describes the interactions supported by the plasmonic nanoslit under consideration \cite{Safari}. This nanostructure acts as a 
 plasmonic tritter by coupling the photonic mode $\hat{b}$ and the two plasmonic modes, described by the operators $\hat{a}$ and $\hat{c}$, to three output modes \cite{Safari}. The photonic mode at the output of the nanoslit is described by $\hat{e}$, whereas the two plasmonic output modes are represented by the operators $\hat{f}$ and $\hat{d}$. As indicated in Fig. \ref{fig:figure1}b, and throughout this paper, we study the conditional detection of the output modes $\hat{d}$ and $\hat{e}$ for a situation in which only the input plasmonic modes of $\hat{a}$ and $\hat{c}$ are excited in the nanostructure. Thus, the photonic mode $\hat{b}$ is assumed to be in a vacuum state. In this case, the plasmonic tritter can be simplified to a two-port device described by the following $2\times 2$ matrix
 
 \begin{equation}
    \left(
    \begin{array}{c}
       \hat{d}\\
       \hat{e}\\
      
    \end{array}
    \right) =  
    \left(
    \begin{array}{cc}
        \kappa& r\\
       \tau&\tau
        \end{array}
    \right )
        \\
     \left(
    \begin{array}{c}
        \hat{a}\\
       \hat{c}
    \end{array}
    \right ).
    \label{eq1}
\end{equation}

The photonic mode $\hat{e}$ is transmitted through the slit and its transmission probability is described by $2\abs{\tau}^2=T_\text{ph}$. Here, $T_\text{ph}$ represents the normalized intensity of the transmitted photons. Moreover, the plasmon-to-plasmon coupling at the output of the nanostructure is given by $\abs{\kappa}^2+\abs{r}^2=T_\text{pl}$. Here, the renormalized transmission (after intereference and considering loss) for the plasmonic fields is described by $T_\text{pl}$. From Fig. \ref{fig:figure1}b, we note that the interference supported by the plasmonic nanoslit shares similarities with those induced by a conventional Mach-Zehnder interferometer (MZI) . More specifically, the two plasmonic modes, $\hat{a}$ and $\hat{c}$, interfere at the location of the nanoslit, which in turn scatters the field to generate the output\cite{biosensordeleon}. The interference conditions are defined by the phase shift induced by the analyte. Plasmonic sensors with nanoslits have been extensively investigated in the classical domain, showing the possibility of ultrasensitive detection using minute amounts of analyte \cite{biosensordeleon,NanoToday, Hwang, Maier, Lee1,Chen:18}.
 
  \begin{figure*}[!ht]
  \centering \includegraphics[width=1\textwidth]{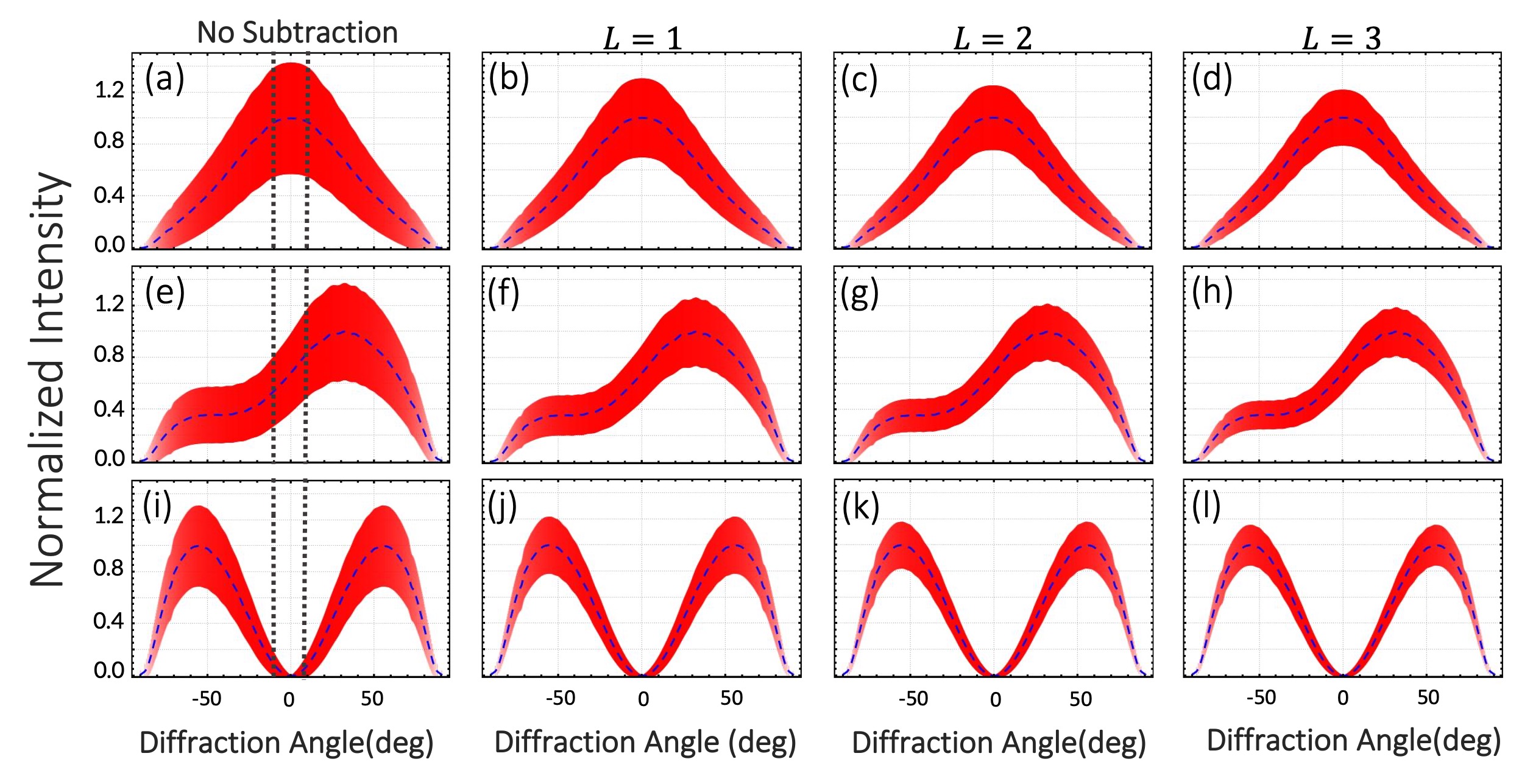}
  \caption{Normalized far-field intensity distribution scattered by the plasmonic nanoslit. The blue dashed line indicates the interference pattern produced by the field transmitted through the 320-nm-wide slit, this corresponds to mode $\hat{e}$. The panels from (a) to (d) are obtained for $\varphi=0$, whereas those from (e) to (h) and (i) to (l) are calculated for $\varphi = \pi/2$ and $\varphi = \pi$ respectively. The dashed line in all plots represents the intensity distribution of the fields transmitted through the slit indicative of dipolar and quadrupolar near-field symmetry for $\varphi=0$ and $\varphi = \pi$. The red shaded regions correspond to the standard deviation for $\bar{n} = 3.75$. Panels (a),(e) and (i) depict the unconditional detection of the signal with its associated noise. As displayed in panels (b) to (d), (f) to (h) and (i) to (l), the signal-to-noise ratio of the plasmonic sensor improves as the fluctuations of the field are reduced through the conditional detection of plasmons. The vertical lines on panels (a), (e) and (i) represent the angular range used for the calculation of the intensity variation with phase (i.e. sensitivity depicted in Fig. \ref{fig:SNRvag2}b ).} 
   \label{fig:figurestdv2}
\end{figure*}

 We now consider a situation in which a single-mode thermal light source  is coupled to the nanostructure in Fig. \ref{fig:figure1}b  exciting two counter-propagating surface plasmon modes. This can be achieved by
using a pair of grating couplers (not shown in the figure) \cite{biosensordeleon}. The statistical properties of this thermal field can be described by the Bose-Einstein statistics as  $
    \rho_{\text{th}} = \sum_{n=0}^{\infty} \text{p}_{\text{pl}}(n) {\ket n} {\bra n}$, where $\text{p}_{\text{pl}}(n)=\bar n^{n}/(1+\bar n)^{1+n}$, and $\bar{n}$ represents the mean occupation number of the field. Interestingly, the super-Poissonian statistics of 
    thermal light can be modified through conditional measurements \cite{Loaiza2019,Dakna,HashemiRafsanjani:17}.
    As discussed below, it is also possible to modify the quantum statistics of plasmonic fields. The control of plasmonic statistics can be implemented by subtracting/adding bosons from/to thermal plasmonic systems \cite{Mizrahi,Parigi}. In this work, we subtract plasmons from the transmitted field formed by the superposition of the surface plasmon modes propagating through the reference
and sensing arms of the interferometer. This is the transmitted mode $\hat{e}$ conditioned on the output of the field $\hat{d}$. The successful subtraction of plasmons boosts the signal of the sensing platforms. This feature is particularly important for sensing schemes relying on dissipative plasmonic platforms.
    
    The conditional subtraction of $L$ plasmons from the mode $\hat{d}$ leads to the modification of the quantum statistics of the plasmonic system, this can be described by
 \begin{equation}
    \text{p}_{\text{pl}}(n) =  \frac{(n+L)!\bar n_{\text{pl}}^{n}}{ n! L!(1+\bar n_{\text{pl}})^{L+1+n}}, 
     \label{eq3prime}
 \end{equation} 
where $\bar{n}_{\text{pl}}$ represents the mean occupation number of the scattered field in mode $\hat{e}$. We quantify the modification of the quantum statistics through the degree of second-order correlation function $g^{(2)}(0)$ for the mode $\hat{e}$ as 
 \begin{equation}
   \begin{split} 
        g^{(2)}_L (0) = {\frac{L+2}{L+1}}.
        \end{split} 
        \label{eqg2}
\end{equation}
We note that the conditional subtraction of plasmons induces anti-thermalization effects that attenuate the fluctuations of the plasmonic thermal system used for sensing. Indeed, the $g^{(2)}_L (0)$ approaches one with the increased number of subtracted plasmons, namely large values of $L$. This effect produces bosonic distributions resembling those of coherent states \cite{maganaloaiza}. Recently, similar anti-thermalization effects have been explored in photonic lattices \cite{Kondakci2015}. 

The aforementioned plasmon subtraction can be implemented in the plasmonic nanoslit interferometer  shown in Fig. \ref{fig:figure1}b. It consists of a 200 nm thick gold film deposited on a glass substrate \cite{biosensordeleon}. This thickness is large enough to enable decoupled plasmonic modes on the top and bottom surfaces of the film, as required.  The gold film features a 320 nm slit, defining the
reference arm of the interferometer to its left and the
sensing arm (holding the analyte) to its right. The analyte then induces
a phase difference $\varphi$  relative to the reference arm, thereby creating the output ($\hat{d}, \hat{e}$ and $\hat{f}$) that depends on this parameter. To verify the feasibility of our conditional measurement approach, we perform a finite-difference time-domain (FDTD) simulation of the plasmonic nanoslit using a wavelength of $\lambda=810$ nm for the two counter-propagating surface plasmon modes ($\hat{a}$ and $\hat{c}$). The nanoslit is designed to support two localized surface plasmon (LSP) modes, one with dipolar symmetry and other with quadrupolar symmetry. Depending on the phase difference $\varphi$, these two LSP modes can be excited with different strengths by the fields interfering at the nanoslit. being the dipolar (quadrupolar) mode optimally excited with $\varphi = 0$ ($\varphi= \pi$). This is due to the fact that the near-field symmetries of the interfering field are well-matched to the dipolar and quadrupolar fields for those values of $\varphi$\cite{biosensordeleon}. The dashed lines in Fig. \ref{fig:figurestdv2} indicate the far-field angular distributions of the transmitted intensity associated with the dipolar LSP mode (panels a to d) and the quadrupolar LSP mode (panels i to l). Only a small angular range of the far-field dis- tribution (range within vertical lines in Fig. \ref{fig:figurestdv2})  is used as the sensing signal. Thus, the sensing signal varies monotonically from a maximum value at $\varphi = 0$  to to a minimum value at $\varphi = \pi$ \cite{biosensordeleon}.

The transmission parameters of our sensor are estimated from FDTD simulations. Specifically, the transmission values for the photonic and plasmonic modes are $T_{\text{ph}}\approx 0.076$ 
and $T_{\text{pl}}\approx0.0176$ 
 for $\varphi= \pi$. However, our subtraction scheme is general and valid for any phase angle  $\varphi$  in the range of $0 \leq \varphi \leq   2 \pi$. Moreover, the total amount of power coupled to modes $\hat{e}$ and $\hat{d}$ normalized to the input power of the plasmonic structure is defined as $\gamma=T_{\text{ph}}+T_{\text{pl}}\approx0.0941$. 
For the results shown in Fig. \ref{fig:figurestdv2}, we assume a mean occupation number of $\bar n=3.75$ for the input beam. As shown in panels (a), (e) and (i) of Fig. \ref{fig:figurestdv2}, the output signals, calculated from Eq. \eqref{eq3prime} and represented by the red shaded region across all panels, exhibit strong quantum fluctuations. Surprisingly, after performing plasmon subtraction, the quantum fluctuations decrease, as indicated in the panels (b)-(d), (f)-(h) and (j)-(l) of Fig. \ref{fig:figurestdv2}. Evidently, this confirms that our conditional measurement protocol can indeed boost the output signal and consequently improve the sensing performance of a plasmonic device. However, due to the probabilistic nature of our protocol and the presence of losses, it is important to estimate the probability rates of successfully performing plasmon subtraction. In Table \ref{tab:table1} we list the degree of second-order correlation $ g^{(2)}_L (0)$, and the probability of successfully subtracting one, two, and three plasmons for different occupation numbers of the plasmonic fields used for sensing. It is worth mentioning that conditional measurements in photonic systems have been experimentally demonstrated with similar efficiencies \cite{Loaiza2019}.
 
 \begin{table}[!htbp]
\centering
\caption{The estimated probability of plasmon subtraction and the corresponding degree of second-order coherence $ g^{(2)}_L (0)$. The losses of the plasmonic nanostructure reduce the probability of subtracting multiple plasmons $L$ from the scattered field with an occupation number of $\bar{n}$. In this case, we assume $\varphi=\pi$.}
\label{tab:table1}
\begin{tabular}{|p{1.5cm}|p{2cm}|p{2cm}|p{2cm}|}
\hline
$\bar{n}$& $ L=1$ &$ L=2 $ &$ L=3$\\
\hline
2&$1.0 \times10^{-2}$ &$1.0 \times10^{-4}$ &$1.1 \times10^{-6}$\\
1&$5.2 \times10^{-3}$ &$2.7\times10^{-5}$ &$1.4 \times10^{-7}$\\
0.5&$2.6 \times10^{-3}$&$7.0 \times10^{-6}$&$1.8 \times10^{-8}$\\
0.3&$1.5\times10^{-3}$&$2.5\times10^{-6} $&$4.0 \times10^{-9} $\\
\hline
$ g^{(2)}_L (0)$&1.5&1.33&1.25\\
\hline
\end{tabular}
\end{table}

The quantities reported in Table \ref{tab:table1} were estimated for a phase shift given by $\varphi=\pi$. This table considers realistic parameters for the losses associated to the propagation of the plasmonic sensing field, and the limited efficiency $\eta_{\text{ph}}$ and $\eta_{\text{pl}}$ of the single-photon detectors used to collect photonic and plasmonic mode respectively. In this case, we assume $\eta_{\text{ph}}=0.3$ and $\eta_{\text{pl}}=0.3$. The latter value is obtained from our simulation, whereas the former corresponds to the efficiency of commercial single-photon detectors \cite{Marsili2013}.  In general, the value for $\varphi$ determines how strongly the dipolar and quadrupolar LSP modes are excited, and consequently their far-field angular distributions. However, the process is applicable for other phases $\varphi$. Our predictions suggest that plasmonic subtraction can be achieved at reasonable rates using a properly designed nanostructure.

\begin{figure}[thp]
  \centering \includegraphics[width=0.85\columnwidth]
  {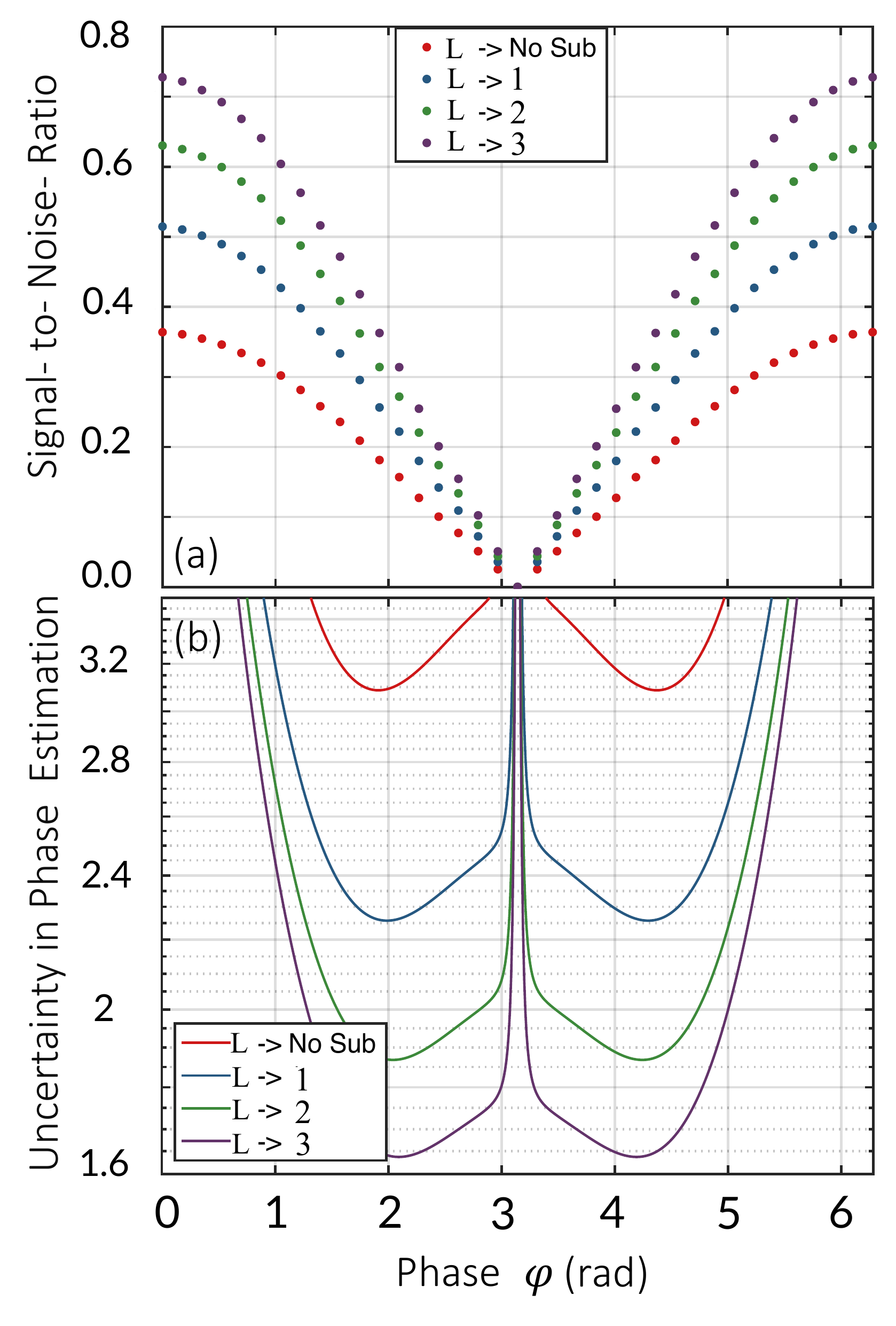}
  \caption{ The panel in (a) reports the signal-to-noise ratio (SNR) as a function of $\varphi$ for the conditional detection of the plasmonic modes transmitted by a 320-nm nanoslit. The red dots represent the unconditional SNR. Furthermore, the blue, green, and purple dots indicate the SNR for the subtraction of one, two, and three plasmons, respectively. This plot shows the possibility of improving the SNR of our plasmonic sensor through the subtraction of plasmons. The panel in (b) indicates that an increasing SNR leads to lower uncertainties in the estimation of phase shifts induced by analytes. The lower uncertainties described by $\Delta\varphi$ imply higher sensitivities of our plasmonic sensor. }
   \label{fig:SNRvag2}
\end{figure}


We now quantify the performance of our conditional scheme for plasmonic sensing through the SNR associated to the estimation of a phase shift. The SNR is estimated as the ratio of the mean occupation number to its standard deviation. This is defined as
\begin{equation}
   \begin{split} 
        \text{SNR} =  \sqrt[]{\frac{(1+L) \bar n \gamma\eta_{\text{ph}} \xi \cos^{2}\frac{\varphi}{2}}{1+ \bar n  \gamma(\xi\eta_{\text{ph}}+(1-\xi)\eta_{\text{pl}}) \cos^{2}\frac{\varphi}{2} }}.
        \end{split} 
        \label{eq4}
\end{equation}

 \noindent 
Here, the parameter $\xi=T_{\text{ph}}/(T_{\text{ph}}+T_{\text{pl}})=0.80 $ 
represents the normalized transmission of the photonic mode. In Fig. \ref{fig:SNRvag2}a, we report the increasing SNR of our plasmonic sensor through the process of plasmon subtraction by plotting the SNR for the subtraction of one, two, and three plasmons for different phase shifts $\varphi$. In addition, for sake of completeness, we evaluate the improvement in sensitivity using error propagation \cite{Hwanglee}. More specifically, we calculate the uncertainty of a phase measurement $\Delta\varphi$. This parameter is estimated as 
 \begin{equation}
   \Delta \varphi=\sqrt{\left\langle \hat{n}^{2}\right\rangle-\langle \hat{n}\rangle^{2}}/\left|\frac{d\langle \hat{n}\rangle}{d \varphi}\right|
        \label{eqg5}
\end{equation}

 Here, the observable $\hat{n}$ corresponds to the conditional intensity measurement within an angular range of the far-field distribution ( specified in Fig \ref{fig:figurestdv2} with the vertical lines). In the field of quantum metrology, the reduced uncertainty of a phase measurement $\Delta\varphi$ is associated to an improvement in the sensitivity of a quantum sensor \cite{Hwanglee,phaseEstimation}. In this regard, the conditional detection of plasmons increases the sensitivity of our plasmonic sensor. This enhancement is reported in Fig. \ref{fig:SNRvag2}b. Here, we demonstrate that the attenuation of the fluctuations of a weak plasmonic field, through the subtraction of up to three plasmons, leads to lower uncertainties in the sensing of photosensitive analytes. 

In conclusion, we have investigated a new method for quantum plasmonic sensing based on the conditional subtraction of plasmons. We have quantified the performance of this scheme, under realistic conditions of loss, by considering the design of a real plasmonic nanoslit sensor. We showed that conditional measurements offer an important path for controlling the statistical fluctuations of plasmonic fields for sensing. In our work, we considered the case for which the sensing field contains a mean plasmonic number lower than two. In this regime, we showed that the attenuation of the quantum fluctuations of plasmonic fields increases the mean occupation number of the sensing field. Interestingly, this effect leads to larger signal-to-noise ratios of our sensing protocol. Furthermore, this feature of our technique enables performing sensitive plasmonic sensing with weak signals \cite{NanoToday, Hwang, Maier, Lee1}.  We believe that our work offers an alternative approach to boost signals in quantum plasmonic platforms operating in the presence of loss at the few particle regime \cite{Tame2013, chenglongreview}. 
\section{FDTD simulation} 
The design of the plasmonic structure given in Fig. \ref{fig:figure1}b is simulated with a 2-D FDTD simulations by a \SI{130}{\micro\meter} domain in $x$ direction and  \SI{8}{\micro\meter} along the $y$ direction. The boundary condition is satisfied via the perfect matching layers to efficiently absorb the light scattred by the strucutre. Besides, the simulations time was long enough so that all energy in the simulation domain was completely decayed. 
The upper clad is made of CYTOP, a polymer with refractive index that closely matches the refractive index of 1.33. The mesh size was as small as $0.03 $ nm 
 along $x$ direction and where we have highly confined field propagation. To create the propagating plasmonic modes, we use a pair of mode sources in both sides of the central slit. The generated SP modes propagate toward the central slit where they interfere. The near-fields along a linear line underneath the nanostrucutre were extracted and used for the far-field analysis. The coupled light to the mode $\hat{e}$, i.e. $T_{\text{ph}}$, was  calculated by the power flow through to the same linear line beneath the slit normalized to the input power.   To have a realistic estimation of the subtracted light, the mode $\hat{d}$  was first propagated for a distance of  10$\lambda$ (8.1 \SI{}{\micro\meter}) along the gold-glass interface and then a grating coupling efficiency of 36\% was considered to out couple the plasmonic mode to the free space  \cite{biosensordeleon}. The out-coupling was done far from the slit to avoid interactions of slit near-fields with fields of the assumed grating.
\section{Acknowledgements}
F.M., C.Y., and O.S.M-L. acknowledge funding from the U.S. Department of Energy, Office of Basic Energy Sciences, Division of Materials Sciences and Engineering under Award DE-SC0021069. M.L.J.L. would like to thank Louisiana State University (LSU) for financial support via the Huel D. Perkins Fellowship, and the LSU Department of Physics \& Astronomy for supplemental support. I.D.L. acknowledges the support of the Federico Baur Endowed Chair in Nanotechnology. 

\section{Competing Interests}
The authors declare no competing interests.

\bibliography{main}

\begin{thebibliography}{10}

\bibitem{NanoToday}
B.~Sepúlveda, P.~C. Angelomé, L.~M. Lechuga, and L.~M. Liz-Marzán,
  ``Lspr-based nanobiosensors,'' {\em Nano Today}, vol.~4, no.~3, pp.~244--251,
  2009.

\bibitem{Hwang}
G.~M. Hwang, L.~Pang, E.~H. Mullen, and Y.~Fainman, ``Plasmonic sensing of
  biological analytes through nanoholes,'' {\em IEEE Sens. J.}, vol.~8, no.~12,
  pp.~2074--2079, 2008.

\bibitem{Maier}
S.~A. Maier, {\em Plasmonics: Fundamentals and Applications}.
\newblock Springer-Verlag GmbH, 2007.

\bibitem{Lee1}
C.~Lee, B.~Lawrie, R.~Pooser, K.-G. Lee, C.~Rockstuhl, and M.~Tame, ``Quantum
  plasmonic sensors,'' {\em Chem. Rev.}, vol.~121, no.~8, pp.~4743--4804, 2021.

\bibitem{ChenglongAIP}
C.~You, M.~Hong, P.~Bierhorst, A.~E. Lita, S.~Glancy, S.~Kolthammer, E.~Knill,
  S.~W. Nam, R.~P. Mirin, O.~S. Magaña-Loaiza, and T.~Gerrits, ``Scalable
  multiphoton quantum metrology with neither pre- nor post-selected
  measurements,'' {\em Appl. Phys. Rev.}, vol.~8, no.~4, p.~041406, 2021.

\bibitem{Slussarenko2017}
S.~Slussarenko, M.~M. Weston, H.~M. Chrzanowski, L.~K. Shalm, V.~B. Verma,
  S.~W. Nam, and G.~J. Pryde, ``Unconditional violation of the shot-noise limit
  in photonic quantum metrology,'' {\em Nat. Photonics}, vol.~11, pp.~700--703,
  Nov 2017.

\bibitem{Polino}
E.~Polino, M.~Valeri, N.~Spagnolo, and F.~Sciarrino, ``Photonic quantum
  metrology,'' {\em AVS Quantum Sci.}, vol.~2, no.~2, p.~024703, 2020.

\bibitem{Altewischer}
E.~Altewischer, M.~P. van Exter, and J.~P. Woerdman, ``Plasmon-assisted
  transmission of entangled photons,'' {\em Nature}, vol.~418, pp.~304--306,
  Jul 2002.

\bibitem{Akimov}
A.~V. Akimov, A.~Mukherjee, C.~L. Yu, D.~E. Chang, A.~S. Zibrov, P.~R. Hemmer,
  H.~Park, and M.~D. Lukin, ``Generation of single optical plasmons in metallic
  nanowires coupled to quantum dots,'' {\em Nature}, vol.~450, pp.~402--406,
  Nov 2007.

\bibitem{Tame2013}
M.~S. Tame, K.~R. McEnery, S.~K. {O}zdemir, J.~Lee, S.~A. Maier, and M.~S. Kim,
  ``Quantum plasmonics,'' {\em Nat. Phys.}, vol.~9, pp.~329--340, Jun 2013.

\bibitem{chenglongreview}
C.~You, A.~C. Nellikka, I.~D. Leon, and O.~S. Magaña-Loaiza, ``Multiparticle
  quantum plasmonics,'' {\em Nanophotonics}, vol.~9, no.~6, pp.~1243--1269,
  2020.

\bibitem{chenglongnature}
C.~You, M.~Hong, N.~Bhusal, J.~Chen, M.~A. Quiroz-Ju{\'a}rez, J.~Fabre,
  F.~Mostafavi, J.~Guo, I.~De~Leon, R.~d.~J. Le{\'o}n-Montiel, and O.~S.
  Maga{\~{n}}a-Loaiza, ``Observation of the modification of quantum statistics
  of plasmonic systems,'' {\em Nat Commun}, vol.~12, p.~5161, Aug 2021.

\bibitem{BenjaminVest}
B.~Vest, M.-C. Dheur, {\'E}.~Devaux, A.~Baron, E.~Rousseau, J.-P. Hugonin,
  J.-J. Greffet, G.~Messin, and F.~Marquier, ``Anti-coalescence of bosons on a
  lossy beam splitter,'' {\em Science}, vol.~356, no.~6345, pp.~1373--1376,
  2017.

\bibitem{Schouten}
H.~F. Schouten, N.~Kuzmin, G.~Dubois, T.~D. Visser, G.~Gbur, P.~F.~A. Alkemade,
  H.~Blok, G.~W.~t. Hooft, D.~Lenstra, and E.~R. Eliel, ``Plasmon-assisted
  two-slit transmission: Young's experiment revisited,'' {\em Phys. Rev.
  Lett.}, vol.~94, p.~053901, Feb 2005.

\bibitem{maganaloaizaexotic2016}
O.~S. Magaña-Loaiza, I.~De~Leon, M.~Mirhosseini, R.~Fickler, A.~Safari,
  U.~Mick, B.~McIntyre, P.~Banzer, B.~Rodenburg, G.~Leuchs, and R.~W. Boyd,
  ``Exotic looped trajectories of photons in three-slit interference,'' {\em
  Nat Commun}, vol.~7, p.~13987, Dec. 2016.

\bibitem{Alexander}
A.~B\"use, M.~L. Juan, N.~Tischler, V.~D'Ambrosio, F.~Sciarrino, L.~Marrucci,
  and G.~Molina-Terriza, ``Symmetry protection of photonic entanglement in the
  interaction with a single nanoaperture,'' {\em Phys. Rev. Lett.}, vol.~121,
  p.~173901, Oct 2018.

\bibitem{Dongfang}
D.~Li and D.~Pacifici, ``Strong amplitude and phase modulation of optical
  spatial coherence with surface plasmon polaritons,'' {\em Sci. Adv.}, vol.~3,
  no.~10, p.~e1700133, 2017.

\bibitem{Vest}
B.~Vest, I.~Shlesinger, M.-C. Dheur, {\'{E}}.~Devaux, J.-J. Greffet, G.~Messin,
  and F.~Marquier, ``Plasmonic interferences of two-particle n00n states,''
  {\em New J. Phys.}, vol.~20, p.~053050, may 2018.

\bibitem{Huck}
A.~Huck, S.~Smolka, P.~Lodahl, A.~S. S\o{}rensen, A.~Boltasseva, J.~Janousek,
  and U.~L. Andersen, ``Demonstration of quadrature-squeezed surface plasmons
  in a gold waveguide,'' {\em Phys. Rev. Lett.}, vol.~102, p.~246802, Jun 2009.

\bibitem{Fasel}
S.~Fasel, F.~Robin, E.~Moreno, D.~Erni, N.~Gisin, and H.~Zbinden, ``Energy-time
  entanglement preservation in plasmon-assisted light transmission,'' {\em
  Phys. Rev. Lett.}, vol.~94, p.~110501, Mar 2005.

\bibitem{Martino}
G.~Di~Martino, Y.~Sonnefraud, S.~Kéna-Cohen, M.~Tame, a.~K. Ozdemir, M.~S.
  Kim, and S.~A. Maier, ``Quantum statistics of surface plasmon polaritons in
  metallic stripe waveguides,'' {\em Nano Lett.}, vol.~12, no.~5,
  pp.~2504--2508, 2012.
\newblock PMID: 22452310.

\bibitem{Heeres2013}
R.~W. Heeres, L.~P. Kouwenhoven, and V.~Zwiller, ``Quantum interference in
  plasmonic circuits,'' {\em Nat. Nanotechnology}, vol.~8, pp.~719--722, Oct
  2013.

\bibitem{Chen:18}
Y.~Chen, C.~Lee, L.~Lu, D.~Liu, Y.-K. Wu, L.-T. Feng, M.~Li, C.~Rockstuhl,
  G.-P. Guo, G.-C. Guo, M.~Tame, and X.-F. Ren, ``Quantum plasmonic n00n state
  in a silver nanowire and its use for quantum sensing,'' {\em Optica}, vol.~5,
  pp.~1229--1235, Oct 2018.

\bibitem{Dowran2018}
M.~Dowran, A.~Kumar, B.~J. Lawrie, R.~C. Pooser, and A.~M. Marino,
  ``Quantum-enhanced plasmonic sensing,'' {\em Optica}, vol.~5, pp.~628--633,
  May 2018.

\bibitem{Lee2016}
C.~Lee, F.~Dieleman, J.~Lee, C.~Rockstuhl, S.~A. Maier, and M.~Tame, ``Quantum
  plasmonic sensing: Beyond the shot-noise and diffraction limit,'' {\em ACS
  Photonics}, vol.~3, pp.~992--999, Jun 2016.

\bibitem{lee}
M.~S. Tame, C.~Lee, J.~Lee, D.~Ballester, M.~Paternostro, A.~V. Zayats, and
  M.~S. Kim, ``Single-photon excitation of surface plasmon polaritons,'' {\em
  Phys. Rev. Lett.}, vol.~101, p.~190504, Nov 2008.

\bibitem{Salazar}
J.~R. Mejía-Salazar and O.~N. Oliveira, ``Plasmonic biosensing,'' {\em Chem.
  Rev.}, vol.~118, no.~20, pp.~10617--10625, 2018.

\bibitem{Kongsuwan}
N.~Kongsuwan, X.~Xiong, P.~Bai, J.-B. You, C.~E. Png, L.~Wu, and O.~Hess,
  ``Quantum plasmonic immunoassay sensing,'' {\em Nano Lett.}, vol.~19, no.~9,
  pp.~5853--5861, 2019.
\newblock PMID: 31356753.

\bibitem{alashnikov}
D.~A. Kalashnikov, Z.~Pan, A.~I. Kuznetsov, and L.~A. Krivitsky, ``Quantum
  spectroscopy of plasmonic nanostructures,'' {\em Phys. Rev. X}, vol.~4,
  p.~011049, Mar 2014.

\bibitem{Mauranyapin2017}
N.~P. Mauranyapin, L.~S. Madsen, M.~A. Taylor, M.~Waleed, and W.~P. Bowen,
  ``Evanescent single-molecule biosensing with quantum-limited precision,''
  {\em Nat. Photonics}, vol.~11, pp.~477--481, Aug 2017.

\bibitem{Dakna}
M.~Dakna, T.~Anhut, T.~Opatrn\'y, L.~Kn\"oll, and D.-G. Welsch, ``Generating
  schr\"odinger-cat-like states by means of conditional measurements on a beam
  splitter,'' {\em Phys. Rev. A}, vol.~55, pp.~3184--3194, Apr 1997.

\bibitem{Loaiza2019}
O.~S. Maga{\~{n}}a-Loaiza, R.~d.~J. Le{\'o}n-Montiel, A.~Perez-Leija, A.~B.
  U'Ren, C.~You, K.~Busch, A.~E. Lita, S.~W. Nam, R.~P. Mirin, and T.~Gerrits,
  ``Multiphoton quantum-state engineering using conditional measurements,''
  {\em npj Quantum Inf.}, vol.~5, p.~80, Sep 2019.

\bibitem{HashemiRafsanjani:17}
S.~M.~H. Rafsanjani, M.~Mirhosseini, O.~S.~M. {n}a Loaiza, B.~T. Gard,
  R.~Birrittella, B.~E. Koltenbah, C.~G. Parazzoli, B.~A. Capron, C.~C. Gerry,
  J.~P. Dowling, and R.~W. Boyd, ``Quantum-enhanced interferometry with weak
  thermal light,'' {\em Optica}, vol.~4, pp.~487--491, Apr 2017.

\bibitem{biosensordeleon}
U.~Felix-Rendon, P.~Berini, and I.~D. Leon, ``Ultrasensitive nanoplasmonic
  biosensor based on interferometric excitation of multipolar plasmonic
  modes,'' {\em Opt. Express}, vol.~29, pp.~17365--17374, May 2021.

\bibitem{Maga_a_Loaiza_2016}
O.~S. Maga{\~{n}}a-Loaiza, J.~Harris, J.~S. Lundeen, and R.~W. Boyd,
  ``Weak-value measurements can outperform conventional measurements,'' {\em
  Phys. Scr.}, vol.~92, p.~023001, dec 2016.

\bibitem{Safari}
A.~Safari, R.~Fickler, E.~Giese, O.~S. Maga\~na Loaiza, R.~W. Boyd, and
  I.~De~Leon, ``Measurement of the photon-plasmon coupling phase shift,'' {\em
  Phys. Rev. Lett.}, vol.~122, p.~133601, Apr 2019.

\bibitem{Mizrahi}
S.~S. Mizrahi and V.~V. Dodonov, ``Creating quanta with an ~annihilation~
  operator,'' {\em J. Phys. A: Math. Gen.}, vol.~35, pp.~8847--8857, oct 2002.

\bibitem{Parigi}
V.~Parigi, A.~Zavatta, M.~Kim, and M.~Bellini, ``Probing quantum commutation
  rules by addition and subtraction of single photons to/from a light field,''
  {\em Science}, vol.~317, no.~5846, pp.~1890--1893, 2007.

\bibitem{maganaloaiza}
O.~S. Magaña-Loaiza, R.~d.~J. León-Montiel, A.~Perez-Leija, A.~B. U’Ren,
  C.~You, K.~Busch, A.~E. Lita, S.~W. Nam, R.~P. Mirin, and T.~Gerrits,
  ``Multiphoton quantum-state engineering using conditional measurements,''
  {\em npj Quantum Inf.}, vol.~5, p.~80, Sept. 2019.

\bibitem{Kondakci2015}
H.~E. Kondakci, A.~F. Abouraddy, and B.~E.~A. Saleh, ``A photonic
  thermalization gap in disordered lattices,'' {\em Nat. Phys.}, vol.~11,
  pp.~930--935, Nov 2015.

\bibitem{Marsili2013}
F.~Marsili, V.~B. Verma, J.~A. Stern, S.~Harrington, A.~E. Lita, T.~Gerrits,
  I.~Vayshenker, B.~Baek, M.~D. Shaw, R.~P. Mirin, and S.~W. Nam, ``Detecting
  single infrared photons with 93{\%} system efficiency,'' {\em Nature
  Photonics}, vol.~7, pp.~210--214, Mar 2013.

\bibitem{Hwanglee}
H.~Lee, P.~Kok, and J.~P. Dowling, ``A quantum rosetta stone for
  interferometry,'' {\em Journal of Modern Optics}, vol.~49, no.~14-15,
  pp.~2325--2338, 2002.

\bibitem{phaseEstimation}
X.-Y. Xu, Y.~Kedem, K.~Sun, L.~Vaidman, C.-F. Li, and G.-C. Guo, ``Phase
  estimation with weak measurement using a white light source,'' {\em Phys.
  Rev. Lett.}, vol.~111, p.~033604, Jul 2013.

\end{thebibliography}
\end{document}